\documentclass[aps,nofootinbib,showpacs]{revtex4}

\begin{document}

\rightline{July 2005}

\title{
Generalized mirror matter models
}

\author{R. Foot}\email{foot@physics.unimelb.edu.au}
\affiliation{School of Physics,
Research Centre for High Energy Physics,
The University of Melbourne, Victoria 3010, Australia}

\begin{abstract}
Non-minimal gauge models with exact unbroken improper space-time
symmetries are constructed and their cosmological and astrophysical
implications explored. 

\end{abstract}
\maketitle

The exact parity model\cite{flv} is the minimal extension of the
standard model which allows for an exact unbroken parity
symmetry [$x \to -x, t \to t$]. Each type of ordinary particle
(lepton, quark, gauge particles) has a distinct mirror
partner.  The ordinary and mirror particles form
parallel sectors each with gauge symmetry $G_{SM} \equiv
SU(3)_c \otimes SU(2) \otimes U(1)_Y$, so that the
gauge symmetry is $G_{SM} \otimes G_{SM}$. The interactions
of each sector are governed
by a Lagrangian of exactly the same form, except with left and right
interchanged. That is, in the mirror sector, it is the right-handed
chiral fermions which are $SU(2)$ doublets, while in the ordinary sector,
it is the left-handed chiral fermions. 
Thus, the Lagrangian has the form:
\begin{eqnarray}
{\cal L} = {\cal L}_{SM} (e_L, e_R, q_L, q_R, W_\mu, A_\mu, ...) 
+ {\cal L}_{SM} (e'_R, e'_L, q'_R, q'_L, W'_\mu, A'_\mu, .... ) + 
{\cal L}_{mix} 
\label{1}
\end{eqnarray}
where $e, q, W_\mu, A_\mu$ denote the leptons, quarks, gauge fields etc
and the primed fields are their corresponding mirror partners.
The ${\cal L}_{mix}$ part describes possible interactions
coupling ordinary and mirror particles together.

The exact
parity symmetry, ${\cal P}$, has the form\cite{flv}:
\begin{eqnarray}
& x \to  -x,  \ t \to t  &\nonumber \\
& G^{\mu} \leftrightarrow G'_{\mu},\  W^{\mu} \leftrightarrow W'_{\mu}, \
B^{\mu} \leftrightarrow B'_{\mu}
&
\nonumber \\
& \ell_{L} \leftrightarrow \gamma_0 \ell'_{R}, \
e_{R}\leftrightarrow \gamma_0 e'_{L}, \
q_{L} \leftrightarrow \gamma_0 q'_{R},  & \nonumber \\
& u_{R} \leftrightarrow  \gamma_0 u'_{L}, \
d_{R} \leftrightarrow \gamma_0 d'_{L}, \
\phi \leftrightarrow \phi'
&
\end{eqnarray}
where $G^{\mu}, W^{\mu}, B^{\mu}$ are the standard $G_{SM}$ gauge
particles, $\ell_{L}, e_{R}, q_{L}, u_{R}, d_{R}$
are the standard leptons and quarks (the generation
index is implicit) $\phi$ is the standard Higgs doublet and 
the primes denote the mirror particles. 
Under this symmetry, ${\cal L}_{SM} (e, q, W, A,...)$
interchanges with ${\cal L}_{SM} (e', q', W', A', ...)$
leaving the full Lagrangian invariant (of course the terms 
in ${\cal L}_{mix}$
must also be invariant under this symmetry). 
The theory also has an exact unbroken time 
reversal invariance, ${\cal T}$,
with standard CPT identified as the product:  ${\cal PT}$\cite{flv}.
In this way, the full Poincare group, containing proper and improper
Lorentz transformations, space-time translations
etc becomes a fundamental unbroken symmetry -- providing
strong theoretical motivation for the theory.

Constraints from gauge invariance and renormalizability limit
${\cal L}_{mix}$ to just two terms\cite{flv}:
\begin{eqnarray}
{\cal L}_{mix} = \epsilon F^{\mu \nu} F'_{\mu \nu} + \lambda 
\phi^{\dagger} \phi \phi'^{\dagger} \phi' 
\end{eqnarray}
where $F^{\mu \nu} = \partial^{\mu} B^{\nu} - \partial^{\nu} B^{\mu}$
[$ F'^{\mu \nu} = \partial^{\mu} B'^{\nu} - \partial^{\nu} B'^{\mu}$]
is the $U(1)$ [mirror $U(1)$] field strength tensor.
The two terms in ${\cal L}_{mix}$ provide an important means of experimentally
testing this theory (for a recent review of these
experimental implications, see Ref.\cite{sad}). 
Note that ${\cal L}_{mix}$ can contain other terms if
there exists new particles, such as gauge singlet neutrinos. In particular
the physics of neutrino mass generation may allow for neutrino - mirror
neutrino mass mixing terms in ${\cal L}_{mix}$, which would 
provide another useful way to test the theory\cite{flv2}.

Note that there is a large
range of parameters of the Higgs potential for which
mirror symmetry is {\it not}
spontaneously broken by the vacuum (i.e. $\langle \phi \rangle =
\langle \phi' \rangle$) so that it is an exact,
unbroken symmetry of the theory. Explicitly, the most general Higgs 
potential is\cite{flv}
\begin{eqnarray}
V = \lambda_1 (\phi^{\dagger} \phi - u^2)^2 + 
\lambda_1 (\phi'^{\dagger} \phi' - u^2)^2 + 
\lambda_2 (\phi^{\dagger}\phi - \phi'^{\dagger}\phi')^2
\label{higgs}
\end{eqnarray} 
If $\lambda_{1,2} > 0$ then
$V \ge 0$. The minimum of the potential, $V = 0$,
occurs when $ \langle \phi \rangle =
\langle \phi' \rangle = u$, demonstrating that the exact parity
symmetry is not broken by the vacuum, as advertised.

Importantly, the theory predicts the existence of 
new particles which are {\it necessarily} massive and stable.
There is a pressing need for such non-baryonic massive stable 
particles from
astrophysical and cosmological considerations, and a significant
amount of work has been done exploring the possibility that
mirror matter is the inferred non-baryonic dark
matter in the Universe\cite{a0, a1, a,b,c,d,e,fm,g,h,i}. 
For an up-to-date review, see ref.\cite{fr}.

The purpose of this article is to examine an obvious generalization
of the exact parity model.
In the minimal exact parity model there is one `copy'
of the standard particles. It is possible that nature may
have $n$ distinct copies of the standard particles. The copies may be mirror
copies or ordinary copies. 

Let us introduce the notation, (p,q) to denote $p$ ordinary sectors
and $q$ mirror sectors. This means that there are $n = p+q-1$ copies
of the standard particles, $q$ of these of the mirror variety. 
We assume an exact parity symmetry, ${\cal P}$,
which implies equal numbers of ordinary and mirror copies: $p=q$
(and hence $n$ is odd).
If the masses and interactions of the particles in each sector are
exactly the same as the standard particles (excepting, of course, that
the mirror copies have left and right interchanged), then
the Lagrangian would exhibit the discrete symmetry 
\begin{eqnarray}
 C_{p}  \otimes C_p \otimes {\cal P} 
\label{dis}
\end{eqnarray}
where $C_p$ is the group of permutations of $p$ objects.

These non-minimal mirror models are a straightforward generalization
to the exact parity model of ref.\cite{flv}.  The Lagrangian 
generalizes Eq.(\ref{1}) in the obvious way:
\begin{eqnarray}
{\cal L} = \sum^p_{i=1}{\cal L}_{SM} (e_{iL}, e_{iR}, q_{iL}, 
q_{iR}, W_i^\mu, A_i^\mu, ...) \ + 
\ \sum^p_{j=1} {\cal L}_{SM}
(e'_{jR}, e'_{jL}, q'_{jR}, q'_{jL},  W'^{\mu}_j A'^{\mu}_j, .... ) 
\ + \ {\cal L}_{mix} 
\label{666}
\end{eqnarray}
where we use the integer subscripts to label the particles from
the $p$ ordinary sectors and primes plus integer subscripts to
label their corresponding mirror partners.
In this general case,
${\cal L}_{mix}$ has the form:
\begin{eqnarray}
{\cal L}_{mix} = \epsilon \sum^p_{i=1} F_i^{\mu \nu} \sum^p_{j=1} 
F'_{j \ \mu \nu} 
\ + \ 
\epsilon' \sum^p_{k,l=1} \left(F^{\mu \nu}_k F_{l\ \mu \nu} \ + \ 
F'^{\mu \nu}_k F'_{l\ \mu \nu}  
\right)
\nonumber \\
\ + \
\lambda \sum^p_{i=1} \phi^{\dagger}_i \phi_i \sum^p_{j=1}
\phi'^{\dagger}_j
\phi'_j
\ + \
\lambda' \sum^p_{k,l=1} \left(\phi^{\dagger}_k \phi_k \phi^{\dagger}_l \phi_l 
\ + \
\phi'^{\dagger}_k \phi'_k \phi'^{\dagger}_l
\phi'_l 
\right)
\end{eqnarray}
where $k \neq l$ in the sums and 
$F_i^{\mu \nu} \equiv \partial^{\mu}B_i^{\nu} - \partial^{\nu}B_i^{\mu}  
\ [F'^{\mu \nu}_i \equiv \partial^{\mu}B'^{\nu}_i -
\partial^{\nu}B'^{\mu}_i  
]$.

The most general Higgs potential is given by the straightforward 
generalization to
Eq.(\ref{higgs}):
\begin{eqnarray}
V &=& \lambda_1 \sum^p_{i=1} \left\{ [\phi_i^{\dagger} \phi_i - u^2]^2 
+ [\phi'^{\dagger}_i \phi'_i - u^2]^2\right\} \ + \
\lambda_2 \sum^p_{i,j=1} [\phi_i^{\dagger}\phi_i -
\phi'^{\dagger}_j\phi'_j]^2 \nonumber \\
& + & \ \lambda_3 \sum^p_{i,j=1} \left\{ [\phi^{\dagger}_i \phi_i - \phi^{\dagger}_j
\phi_j]^2
+ [\phi'^{\dagger}_i \phi'_i - \phi'^{\dagger}_j
\phi'_j]^2
\right\}
\end{eqnarray}
Again, if $\lambda_{1,2,3} > 0$ then 
$V \ge 0$. The minimum of the Higgs potential is
then  $V=0$, which occurs only when each $\langle \phi_i \rangle
= \langle \phi'_i \rangle
= u$. This means that the discrete symmetry, Eq.(\ref{dis}),
is {\it not} spontaneously broken, but is an exact
unbroken symmetry of the theory.

One specific motivation for considering such non-minimal
models comes from the similarity of the cosmological mass density of
non-baryonic dark matter and ordinary matter.
Precision measurements of the CMBR from WMAP
and other data give\cite{cosmo}:
\begin{eqnarray}
\Omega_b h^2 = 0.0224 \pm 0.0009 \nonumber \\
\Omega_m h^2 = 0.135^{+0.008}_{-0.009}
\end{eqnarray}
where $\Omega_m = \Omega_b + \Omega_{dark}$ is the total matter
density (normalized to the critical matter
density needed to close the Universe) and $h$ is the Hubble
parameter measured in units of $100$ km/s/Mpc.
This means that the cosmological mass density of non-baryonic dark matter
is within an order of magnitude of the mass density of baryons:
\begin{eqnarray}
\Omega_{dark}/\Omega_b = 5.03 \pm 0.46
\ .
\end{eqnarray}
This interesting result can be explained in principle if the 
mass and interaction rates of the non-baryonic particles
are very similar to ordinary baryons. 
The exact parity model
is one simple and well defined extension of the standard model
which has this feature\cite{mon,mon2}.
In fact, in the minimal exact parity model, 
we would expect $\Omega_{dark} =
\Omega_b$ if the evolution of the universe were completely 
symmetric in the two sectors, i.e. there was no temperature difference 
between the ordinary and mirror particles
during the baryogenesis 
epoch\footnote{Actually, $\Omega_{dark} = \Omega_b$
could also occur -- even if there was a temperature asymmetry
during baryogenesis -- if the baryonic
asymmetry was generated by transitions between ordinary and mirror
particles as in the scenario of Ref.\cite{mon}.}.
During the big bang nucleosynthesis (BBN) epoch, 
however, the success of standard BBN
suggests that $T'$ is less than $T$:
\begin{eqnarray}
T'/T \stackrel{<}{\sim} 0.6 \ {\rm at} \
T \sim 1 \ {\rm MeV}
\label{bbn}
\end{eqnarray}
in order for the expansion rate of the universe to have been
within an acceptable range.
If this temperature asymmetry were induced\footnote{
The origin of the temperature asymmetry is unknown, however some
ideas have been put forward in Ref.\cite{inf} in the
context of inflation.  
The basic idea is to have an `ordinary inflaton' coupling
to ordinary matter, and a `mirror inflaton' coupling
to mirror matter. If inflation is triggered by some random fluctuation,
then it can occur in the two sectors at different times,
leading to $T \neq T'$ after reheating in the two sectors.
Naturally, the bulk of the inflation would be expected
to occur prior to baryogenesis (so as not to dilute the
baryon number), however this does not exclude the
possibility of asymmetric reheating after baryogenesis, since
the Universe may have gone through several reheating processes - depending
on details such as the
the number of weakly coupled scalar fields.}
before baryogenesis, then
we expect $\Omega_{dark} \neq \Omega_b$, the details
would, of course, depend on the precise model of
baryogenesis. Obviously if the
temperatures of the two sectors were not too different this
might explain why $\Omega_b \sim \Omega_{dark}$ (see also Ref.\cite{mon,mon2} for
some related scenarios). 
Another logical possibility is that the temperature asymmetry required
by BBN was 
induced after baryogenesis. In this case, the abundance of
mirror baryons would be exactly the same as the abundance of
ordinary baryons, and in the non-minimal mirror models with $n$ copies, this
would generalize to:
\begin{eqnarray}
\Omega_{dark}/\Omega_b = n
\ .
\end{eqnarray} 
This specific scenario is obviously testable and falsifiable, 
since it predicts that this
ratio is an odd integer. The data are currently consistent with that, 
suggesting $n=5$. 
Thus we are led to consider the specific particle
physics model consisting of the ordinary particles and 5
copies.
This is compatible with our hypothesis of exact parity symmetry,
implying three ordinary and three mirror sectors.

The successful dark matter features of the mirror matter model would also
occur in these non-minimal models, including:
\begin{itemize}

\item
It would elegantly explain\cite{a0} 
the MACHO population inferred to exist in the
galactic halo from numerous microlensing observations\cite{ml} of nearby
galaxies.

\item
It would be capable of explaining the large scale structure
of the universe\cite{a,b,c}.

\item
Provide a straightforward explanation\cite{footdama} of the 
positive annual modulation signal obtained in the
DAMA/NaI direct detection experiment\cite{dama}. Importantly, this
explanation is consistent with the null results
of the other direct detection experiments\cite{footdama2,sad}.

\item
Explain various solar system anomalies, such as the anomalous 
acceleration\cite{pioneer} of the two Pioneer spacecraft\cite{fvp}, 
lack of small craters on the asteroid 433 Eros \cite{fm} etc.

\end{itemize}

If mirror matter is the non-baryonic dark matter, as the 
above experiments and observations suggest, then the dark halos
inferred to exist in spiral galaxies should be composed 
predominately of mirror matter.
Microlensing studies\cite{ml} suggest a 
halo composed of about $20\%$ mirror stars
with the rest presumably in the form of an ionized mirror gas.
Ionized gas, with a typical virial temperature of order $T \sim 100$ eV
radiates energy at a rate per unit volume of\cite{book}:
\begin{eqnarray}
r_{cool} = n^2_{e'} \Lambda
\end{eqnarray}
where $n_{e'}$ is the (free) mirror electron number density
and $\Lambda$ is a calculable function (which depends on cross section,
temperature, composition etc). For a temperature of $T \sim 100 \ eV,
\Lambda \sim 10^{-23}\ erg \ cm^3 \ s^{-1}$ (see e.g. Ref.\cite{book}).
In the case of the minimal mirror model, i.e. with $n=1$, 
the halo would have a mirror photon luminosity of\cite{e}:
\begin{eqnarray}
L_{halo} = \int^{100 kpc}_{R_1} n_{e'}^2 \Lambda 4\pi r^2 dr
\sim \left( {3 kpc \over R_1}\right) 10^{44}\ erg/s.
\end{eqnarray}
where $R_1$ is a phenomenological cutoff.
If there are $n$ copies, and the particles in each sector
are equally abundant
in the halo of the galaxy, then $L_{halo}$ is reduced by $n^2$ for
each type of photon.
Thus, for $n=5$, the luminosity in each type of photon is
only about a few times $10^{42}$ erg/s. The heating
responsible for supporting the halo is not completely
clear, however plausible candidates include mirror and/or ordinary 
supernova which can potentially supply the
required energy\cite{e}.

Obviously, since the ordinary particles collapse and form a disk, while
the mirror particles are roughly spherically distributed 
in spiral galaxies
\footnote{
Here, we use the term `mirror particles' as an inclusive
term for particles of the n-copies, whether they are mirror copies or
not.}, 
the evolution is clearly asymmetric. 
Such an asymmetric evolution is plausible 
because the ordinary and 
mirror particles had different temperatures at the epoch of BBN,
Eq.(\ref{bbn})\footnote{
Actually in the case of 5-copies, the BBN bound is slightly more stringent,
$T'/T \stackrel{<}{\sim} 0.4$, assuming a common temperature, $T'$ for 
each of the 5 copies.}.
This temperature asymmetry, not only leads to successful large
scale structure formation\cite{a,b,c}, but also 
implies that the chemical composition of the mirror worlds
are quite different to the ordinary particle world\cite{a}.
Specifically, the proportion of mirror helium ($He'$)
to mirror hydrogen ($H'$) in each of the mirror sectors is
expected to be significantly greater
than the corresponding ordinary $He/H$ ratio\cite{a}.  
Because of this major difference,
the macroscopic evolution of the ordinary sector
will be quite different to that of the mirror sectors. For example, the higher
$He'/H'$ ratio implies that mirror stars evolve much faster (an
order of magnitude or more) than
ordinary stars\cite{brecent}, potentially giving a much greater rate
of mirror supernova explosions. Evidently, 
the ordinary-mirror particle
asymmetry required to explain a) BBN, b) Large scale structure
formation, and c) the disparate 
distribution of ordinary and mirror particles within spiral galaxies
might all be the result of an effective asymmetric boundary
condition, 
with the microscopic interactions, as defined in the
quantum field theoretic Lagrangian, remaining completely symmetric. 
Of course, understanding the complete details of galaxy 
formation (especially in the non-linear regime) is far from
begin fully understood.

In conclusion, we have examined a straightforward generalization of the
exact parity model involving $n$ copies of the standard particles (with
$n$ odd, 
if the fundamental interactions respect an exact parity
symmetry).
Successful early universe cosmology requires a temperature asymmetry
between the
ordinary particles and their $n$ copies at the BBN epoch.
Under the assumption that this temperature asymmetry
arose after baryogenesis, the inferred non-baryonic
dark matter density $\Omega_{dark}/\Omega_b \approx 5$ suggests
that $n = 5$. 
Although non-minimal, such models do preserve the
successful dark matter features inherent in the minimal exact parity
symmetric model and the additional prediction that $\Omega_{dark}/\Omega_b$ is
an odd integer can be further tested by future cosmological observations.

\vskip 0.5cm
\noindent
{\bf Acknowledgements}
\vskip 0.3cm
\noindent
This work was supported by the Australian Research Council.
The author would like to thank H. Georgi for pointing out
some deficiencies in a previous version of this paper.

\end{document}